\documentclass[aps,prl,reprint]{revtex4-1}
\usepackage[usenames,dvipsnames]{color}
\usepackage{amsmath,array,amssymb}
\usepackage{graphicx}
\usepackage{hyperref}
\usepackage{textcomp}
\usepackage{braket}
\usepackage{multirow}
\usepackage{indentfirst}
\usepackage[FIGTOPCAP,raggedright,nooneline,bf,footnotesize]{subfigure}
\usepackage{paralist}
\usepackage{overpic}
\usepackage{gensymb}

\renewcommand{\eqref}[1]{Eq.~(\ref{#1})}
\newcommand{\tbyag}{Tb$^{3+}$:Y$_3$Al$_5$O$_{12}$}

\begin{document}
\title{Modification of relaxation dynamics in Tb$^{3+}$:Y$_3$Al$_5$O$_{12}$ nanopowders}
\author{Thomas Lutz$^1$, Lucile Veissier$^{1*}$, Philip J. T. Woodburn$^2$, Rufus L. Cone$^2$, Paul E. Barclay $^1$, Wolfgang Tittel$^1$, Charles W. Thiel$^2$}
\address{
$^1$Institute for Quantum Science and Technology, and Department of Physics \& Astronomy, University of Calgary, Calgary Alberta T2N 1N4, Canada
\\
$^2$Department of Physics, Montana State University, Bozeman, MT 59717 USA\\
$^*$Present address:  Laboratoire Aim\'e Cotton, CNRS-UPR
3321, Univ. Paris-Sud, B\^at. 505, F-91405 Orsay Cedex,
France 
}

\begin{abstract}
Nanostructured rare-earth-ion doped materials are increasingly being investigated for on-chip implementations of quantum information processing protocols as well as commercial applications such as fluorescent lighting. However, achieving high-quality and optimized materials at the nanoscale is still challenging. Here we present a detailed study of the restriction of phonon processes in the transition from bulk crystals to small ($\le$ 40 nm) nanocrystals by observing the relaxation dynamics between crystal-field levels of Tb$^{3+}$:Y$_3$Al$_5$O$_{12}$. We find that population relaxation dynamics are modified as the particle size is reduced, consistent with our simulations of inhibited relaxation through a modified vibrational density of states and hence modified phonon emission. However, our experiments also indicate that non-radiative processes not driven by phonons are also present in the smaller particles, causing transitions and rapid thermalization between the levels on a timescale of $<$100 ns. 
\end{abstract}

\maketitle
\newpage

\section{Introduction}
Crystalline materials doped with impurities such as rare-earth-ions (REI), or diamond silicon-vancancy (SiV) and nitrogen-vacancy (NV) centers, have found many applications in fields as diverse as quantum information processing \cite{Riedmatten2015,heshami_quantum_2016,Sipahigil2016,Hensen2015}, quantum memories \cite{Hedges2010,saglamyurekbroadband2011,Maurer1283}, sensing \cite{hong2013}, lasers \cite{Powell98}, and phosphors \cite{kenyon_recent_2002,Justel98}. Nanometer-sized structures fabricated from these materials have begun to be investigated for on-chip implementations of these applications. In addition, small-sized nano-phosphors are desired for high-quality window materials used in lamps as well as for state-of-the-art displays \cite{Wang10,Downing1185}. Finally, nano-powders have also been proposed for optical refrigeration applications where their modified phonon spectrum and particle morphology could enhance the cooling efficiency \cite{Ruan2006}.

Nano-materials can be obtained through different routes: nano-structures can be milled or etched from high-quality bulk materials, and nanocrystals can be obtained through chemical synthesis, mechanical crushing, or ablation techniques. The transition to nano-sized structures generally introduces detrimental effects such as poor crystal quality, surface effects, and amorphous behavior that can severely restrict practical applications \cite{coprec,Lutz2017}. While some of these effects, such as the increasing surface-to-volume ratio, are fundamental, others can be minimized by optimizing the fabrication process \cite{coprec}. Indeed, in some cases, both chemical synthesis as well as fabrication methods starting with bulk materials produced high-quality materials \cite{Zhong2015,Ferrier2013}. However, none of those structures have allowed studying the effects of decreasing dimensions on phonon-mediated population dynamics. Furthermore, a general procedure for achieving consistently high-quality nano-materials is still unknown and many open questions remain regarding the transition to smaller sizes, requiring further systematic studies.

During the transition from a bulk material to nano-structures, many material properties change. One interesting effect is the predicted modification of the vibrational density of states (VDOS) in small structures. Whereas a bulk crystal has a Debye VDOS (a continuous function that increases with the square of the vibrational frequency), the distribution becomes discrete in small crystallites, exhibiting  gaps and even a cutoff below which no phonons are supported. Furthermore, phononic crystals---nano-machined, periodic structures---can feature engineered frequency band-gaps where vibrations are forbidden \cite{Lutz2016}. These approaches to phonon engineering could potentially benefit applications in the field of quantum information, in particular quantum memories, since the absence of phonons could enhance spin-population lifetimes as well as optical coherence times. Modifications of population dynamics in REI-doped nanocrystals have been previously reported by Meltzer \textit{et\,al.}\ \cite{Yang1999,Yang1999a}, Liu \textit{et\,al.}\ \cite{Liu2002} and Mercier \textit{et\,al.}\ \cite{Mercier2006}, and it was suggested that the changes were due to phonon suppression in the nanocrystals. However, the particles employed in some of those studies were not sufficiently small to suppress phonons at the desired frequencies, and locally elevated temperatures caused by the optical excitation of the powder materials might explain some of the observed effects. Thus, unambiguous confirmation of the suppression of phonon-mediated relaxation in optical centers through VDOS-engineering remains an open challenge.

In this manuscript we examine the effect of the transition from bulk crystals to $\le$ 40 nm particles (see Fig.~\ref{fig:tb_yag_levels}) on the population dynamics between  excited-state crystal-field levels in Tb$^{3+}$-doped Y$_3$Al$_5$O$_{12}$ (YAG). Specifically, we study the influence of size restriction on relaxation dynamics and equilibrium population distribution between the crystal-field levels, i.e. thermalization. We observe that the population dynamics are strongly modified for smaller particles, which can be explained by a modified density of states. However, we also find that the thermal population distribution exhibited by the nano-powders is the same as in the bulk material. As described in detail in the discussion in later sections, the observation of rapid thermalization suggests that in addition to possible phonon suppression, other non-phononic processes---e.g. related to surface defects or energy transfer \cite{Forster}---are introduced, enabling rapid thermalization of population between the closely-spaced energy levels. 

\section{Experimental Details}
\begin{figure}[t]
\centering
\includegraphics[width=0.75\columnwidth]{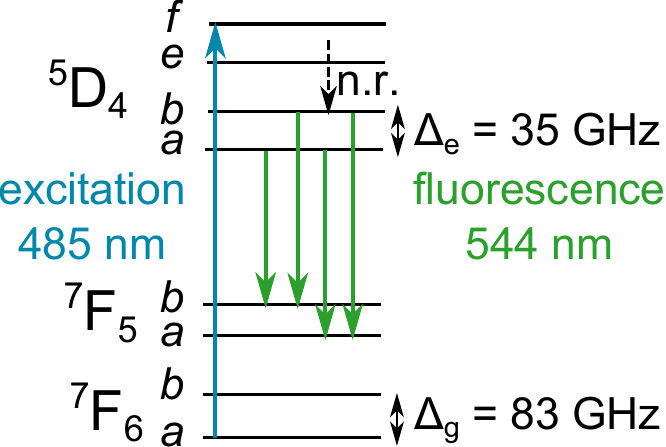}
\caption{Relevant energy levels of \tbyag (vertical axis not to scale) for the measurement of population relaxation between the first two crystal-field levels within the $^5$D$_4$ excited-state manifold. A pulsed laser excites the ions from the $^7$F$_6 \, a$ ground state to the $^5$D$_4 \, f$ excited state, from where they decay rapidly into $^5$D$_4 \, a$ and $b$. The resulting fluorescence due to the four $^5$D$_4 \, a/b \rightarrow ^7$F$_5 \, a/b$ transitions are collected, spectrally resolved, and then analyzed.}
\label{fig:tb_yag_levels}
\end{figure}

We chose Tb$^{3+}$ doped Y$_3$Al$_5$O$_{12}$ (YAG) since the combination of its energy level structure and its high acoustic velocity is well suited to investigate the effects of size on the direct phonon process. More precisely, the small excited-state splitting $\Delta_e=35$~GHz between the Tb$^{3+}$ crystal-field levels $^5$D$_4 \, a/b$ (see Fig.~\ref{fig:tb_yag_levels}), together with the acoustic velocity of 6400  m/s \cite{mezeix_comparison_2006}, results in an expected suppression of the direct phonon process for relatively large particles of $\sim100$ nm diameter. In addition, in the bulk crystal, the inhomogeneous broadening of about 20~GHz allows one to selectively address each of the $^5$D$_4 \, a/b$ crystal-field levels. Furthermore, the ground state splitting $\Delta_g=83$~GHz is large enough that resonant phonons are not expected to be inhibited in the $> 40$~nm diameter nanocrystals, and we can therefore directly measure the internal sample temperature through the ratio of population in the two levels $^7$F$_6\, a/b$.

We created our powders using a sol-gel synthesis (method 1 \cite{kaithwas}) and a modified sol-gel synthesis with a freeze drying step under vacuum and at temperatures below -20~C to restrict agglomeration (method 2 \cite{freeze_dry}). Each method leads to slightly different particle morphologies and size distributions. Additional size control can be achieved by changing the annealing duration and temperature. In this way we were able to vary the nanocrystal diameter $d$ between 40 and 500 nm, and 40 and 70 nm using method 1 and 2, respectively. We evaluated the crystal quality of our powders using a scanning electron microscope, a transmission electron microscope, x-ray diffraction, and optical spectroscopy methods (APPENDIX A and B). From these measurements we conclude that the bulk crystal quality can be maintained for crystallites with diameters down to $\sim 80$~nm. For smaller sizes, we observe a slight decrease in crystal quality that manifests itself in an increase of the inhomogeneous broadening. Measurements of the radiative lifetime as a function of particle size (APPENDIX A) confirm that we can approximately treat the powders as individual particles rather than large agglomerates, with method 2 producing less agglomeration than method 1.

All powders were mounted in an unsealed glass cuvette within an Oxford Instruments cryostat. The samples were held at temperatures from $\sim$  1 K up to 10 K; for temperatures below 2.17 K, the samples were immersed in superfluid liquid helium, whereas for higher temperatures, the samples were cooled by a continuous flow of helium vapor. A pulsed H\"ansch-style nitrogen-laser-pumped dye laser \cite{Haensch1972} with peak powers of up to 10 kW, a pulse duration of 6 ns and a repetition rate of 6 Hz was used with Coumarin 481 dye to provide excitation light at 485 nm. As shown in Fig. \ref{fig:tb_yag_levels}, we resonantly excited Tb ions from the ground state $^7$F$_6\, a$ to the $^5$D$_4 \, f$ \ level, a transition that provides strong absorption. From the  $^5$D$_4\, f$ \ level, the population rapidly decays ( $<$  ns) non-radiatively by emission of high-frequency acoustic phonons into the two levels $^5$D$_4\, a$ \ and $b$. Using a SPEX 1401 monochromator ($<$  3 GHz resolution), we selectively collected fluorescence from each of the four  $^5$D$_4\, a/b$  $\rightarrow$ $^7$F$_5\, a/b$ transitions. The collection was at an angle of 90$\degree$ relative to the excitation laser, 
and the fluorescence was measured using a photomultiplier tube (Hamamatsu R928) terminated with a variable resistance that allowed for time resolutions as fast as 100 ns.

For all powders and experimental configurations described below, we directly measured the local temperature through the relative population in the two levels $^7$F$_6\, a$ \ and $b$, as detailed in APPENDIX C. We found no local heating, thus confirming that observed changes in relaxation dynamics were not due to elevated sample temperatures.

\section{Time-Resolved Fluorescence Measurements}
\begin{figure}[]
\centering
\includegraphics[width=1\columnwidth]{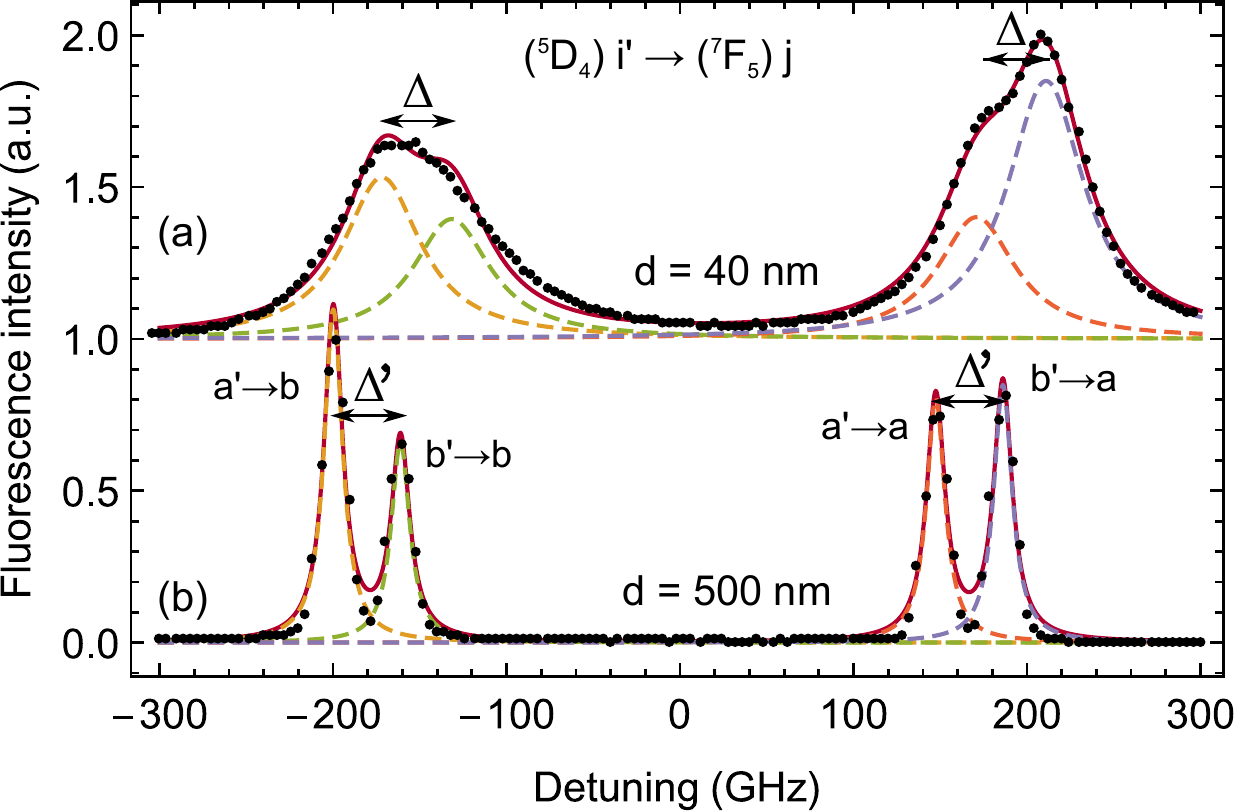}
\caption{Fluorescence spectra at 5 K of (a) a 40 nm diameter powder and (b) a 500 nm diameter powder (synthesized via method 1). Each spectrum (black dots) is fit with the sum of four identical Lorentzian lines with the same pairwise energy splitting $\Delta$ ($\Delta^{\textrm{'}}$). For $d=40$ nm the width is 53 GHz and $\Delta$ is 40 GHz, and for $d=500$ nm the width is 13 GHz and $\Delta^{\textrm{'}}$ is 39 GHz.}
\label{fig:fluospect}
\end{figure}

Before studying the population dynamics of the $^5$D$_4 \, a/b$ \ levels, we first ensured
that we could selectively collect fluorescence from the two excited levels $^5$D$_4 \, a/b$ for each of our samples. We recorded fluorescence spectra by scanning the monochromator over the four lines connecting $^5$D$_4 \, a/b \rightarrow ^7$F$_5 \, a/b$, with typical fluorescence spectra shown in Fig.~\ref{fig:fluospect}. We observed that the smallest nanocrystals  feature an increased inhomogeneous broadening compared to the bulk (for details see APPENDIX B). As a consequence, some fraction of the detected emission originates from the neighboring transition. Since this can lead to observations that could wrongfully be interpreted as modifications in relaxation dynamics, it must be considered in the analysis of any obtained data.

Following this initial characterization, we recorded fluorescence decays from the two crystal-field levels $^5{\rm D}_4 \, a/b$ at a temperature of 1.5~K.  The individual decays were collected by sequentially tuning the monochromator on resonance with each of the four transitions $^5{\rm D}_4\ \, a/b \rightarrow \, ^7{\rm F}_5\, a/b$. The specific frequencies of these transitions were obtained from the fluorescence spectra measured for each sample as described above. 
After excitation of the $^5{\rm D}_4 \, f\;$ level,  2 THz above $^5{\rm D}_4 \, a$, the population first decays rapidly into both $^5{\rm D}_4 \, a/b$ levels and from there into $^7{\rm F}_5 \, a/b$. The dynamics are captured by the following rate equations:

\begin{figure}[t!]
\centering
\includegraphics[width=1\columnwidth]{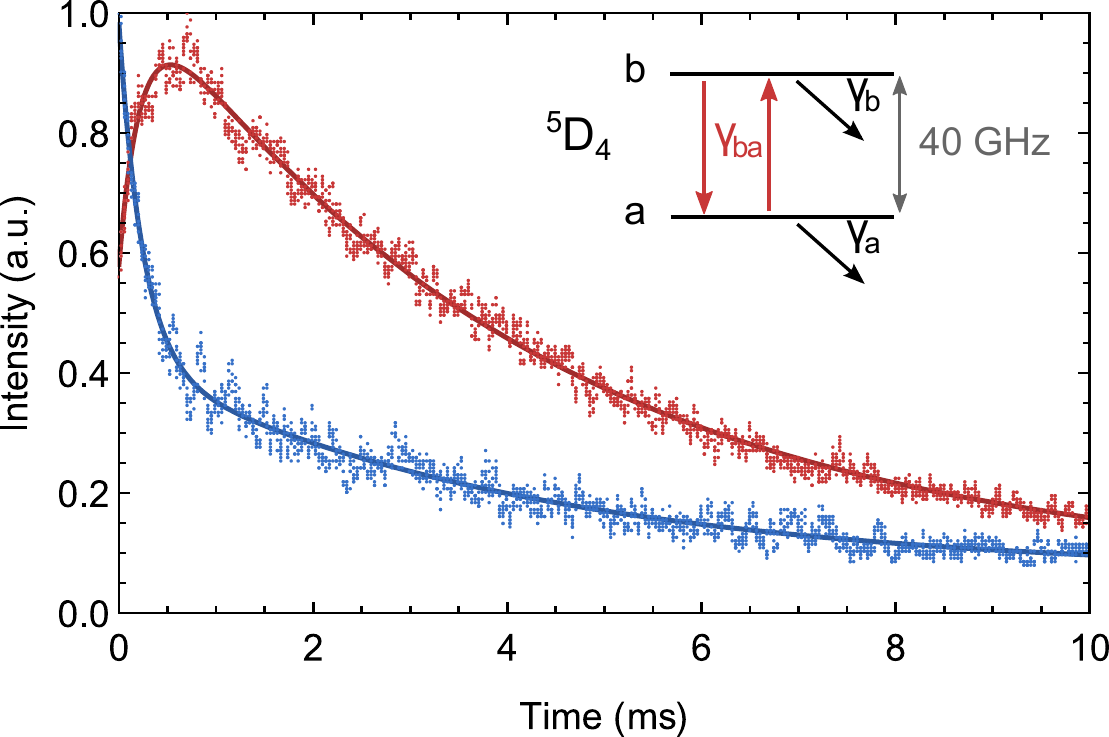}
\caption{Fluorescence decays $^5{\rm D}_4 \, a$ $\rightarrow$ $^7{\rm F}_5 \, a$ (red dots) and $^5{\rm D}_4 \, b$ $\rightarrow$ $^7{\rm F}_5 \, a$ (blue dots) at 1.5 K in large crystallites ($d=500$~nm). The experimental points are fit with Eq.~\ref{eq:2expdecay} (solid lines), resulting in $T_\alpha=$ 235 $\mu$s and  $T_\beta = $ 4 ms. Inset: relevant level structure and rates associated with the fluorescence decays (see definitions in the main text).}
\label{fig:decays}
\end{figure}

\begin{align}
\dot{n}_a(t) &= \gamma_{ba} \left[ n_b(t) - n_a(t) \right] - \gamma_a n_a(t) \\
\dot{n}_b(t) &= - \, \gamma_{ba} \left[ n_b(t) - n_a(t) \right] - \gamma_b n_b(t)
\label{rateequations}
\end{align}
where $n_{a/b}(t)$ denote the populations in the levels $^5{\rm D}_4 \, a/b$ at a time $t$ after excitation, $\gamma_{a/b}$ are the rates of the radiative decay from each level into the $^7{\rm F}_5$ multiplet, and $\gamma_{ba}$ is the rate of the non-radiative process coupling the levels $^5{\rm D}_4 \, a/b$. Note that relaxation into the ground state multiplet is ignored (experimentally and in Eq. \ref{rateequations}) due to the transitions' comparably small rates.

For all four transitions (see Fig.~\ref{fig:decays} for two examples), we observed fluorescence decays composed of two components. The first component corresponds to the non-radiative decay of population from $^5{\rm D}_4 \, b$ to $^5{\rm D}_4 \, a$ and manifests in the fluorescence decays from $^5{\rm D}_4 \, a$ to $^7{\rm F}_5 \, a/b$ as a fill-in, i.e. an increase of the fluorescence intensity with time, and in the fluorescence from $^5{\rm D}_4 \, b$ to $^7{\rm F}_5 \, a/b$ as an initial, fast decay, both with the same characteristic time of about 0.2~ms. 
The second, long component corresponds to the radiative decay. Thus, we fit the recorded, time-dependent fluorescence intensities $I_{a,b}(t)$ from the levels $^5{\rm D}_4 \, a,b$ using
\begin{equation}
I_{a,b}(t) = \tilde{\alpha}_{a,b} e^{-t/T_{\alpha}} + \beta_{a,b} e^{-t/T_{\beta}} ,  
\label{eq:2expdecay}
\end{equation}
where $T_\alpha$ and $T_\beta$ are the time constants of the fast (non-radiative) and slow (radiative) components with corresponding amplitudes $\tilde{\alpha}_{a,b}$ and $\beta_{a,b}$. As discussed before, due to increased inhomogeneous broadening in the small powders, the fluorescence signals collected from transitions starting in either $^5{\rm D}_4 \, a$ or $b$ \ contain a certain amount of emission from the other transition. Fitting our fluorescence spectra with four Lorentzians allows us to compute the percentage $P_{a,b}$ of the collected signal that originates from the neighboring transition.   This affects the recorded signal since the fill-in and the fast decay originating from $^5{\rm D}_4 \, a$ and $b$, respectively, compensate each other to a certain degree. However, the time constants and the amplitudes of the slow decays are not affected. After obtaining $P_{a,b}$ from our fits, we subsequently compute the corrected amplitudes $\alpha_{a,b}=\tilde{\alpha}_{a,b}/(1-2P_{a,b})$. Since we obtained similar results from decays to the levels $^7{\rm F}_5 \, a$ and $b$, the following observations correspond to averages over all four transitions unless otherwise stated. 

\begin{figure}[t!]
\centering
\includegraphics[width=1\columnwidth]{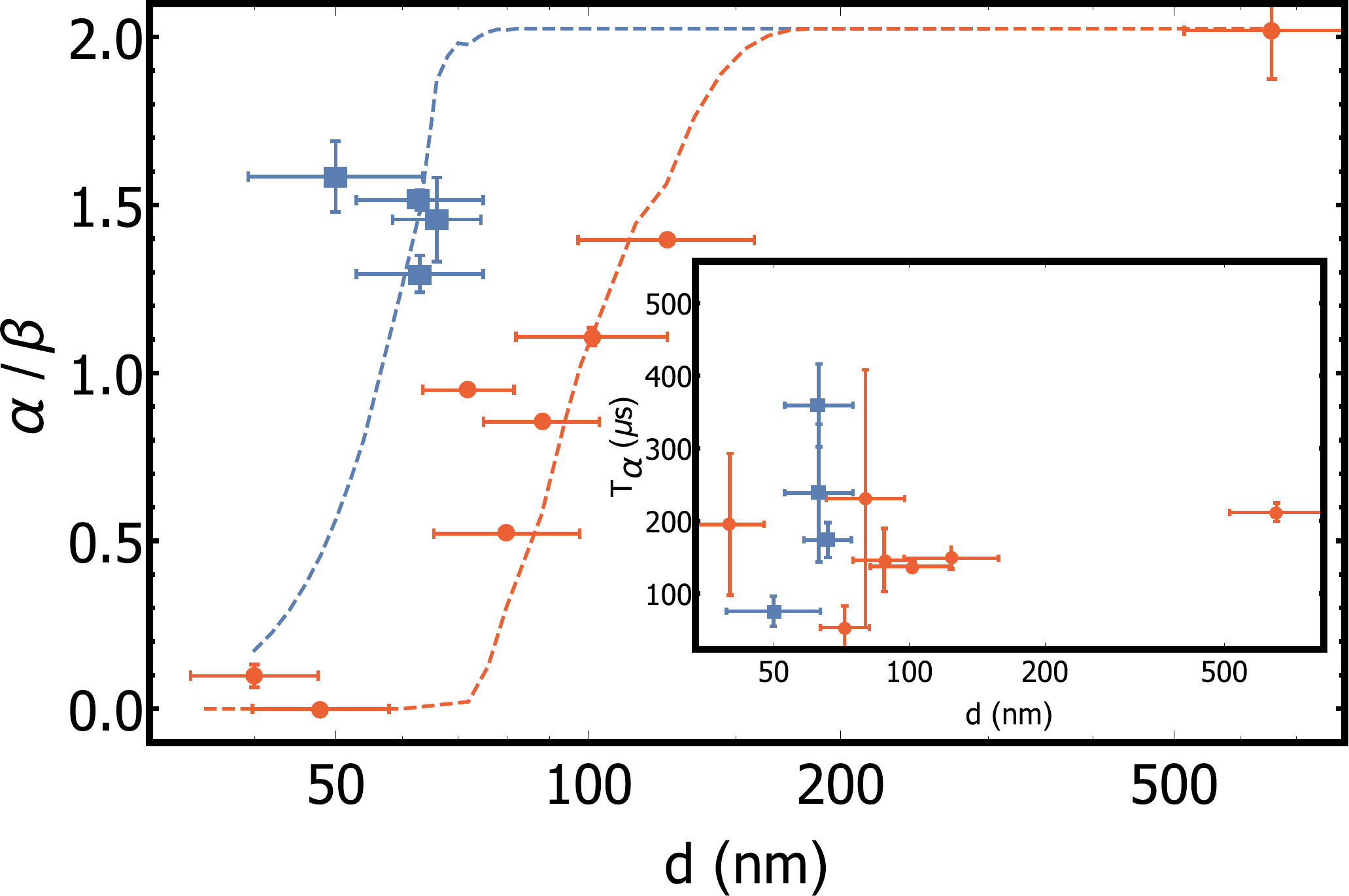}
\caption{Amplitude ratio $\alpha / \beta$ obtained from an average of the fits of Eq.~\ref{eq:2expdecay} to the four decays $^5{\rm D}_4 \, a,b$ $\rightarrow$ $^7{\rm F}_5 \, a,b$ as a function of average nanocrystal diameter $d$ for powder 1 (red circles) and powder 2 (blue squares) at 1.5 K. The horizontal error bars correspond to one standard deviation of the size distribution of the respective sample, obtained through SEM analysis. Simulations are depicted using dashed lines. As described in the main text, the difference between the two simulations is the width of the phonon modes that can be different due to the different morphologies of the two synthesis methods. Note that the amplitude of the fast decay has been corrected for the partially overlapping lines as described in the main text. Inset: Characteristic time of the fast decay component as a function of nanocrystal diameter $d$}
\label{fig:Anr_Tb_vs_size}
\end{figure}

By reducing the size of the particles, we expect to introduce a frequency cutoff in the VDOS below which phonons are not allowed. Consequently, we expect to observe a change in the non-radiative relaxation between the two closely spaced crystal-field levels $^5{\rm D}_4 \, a$ and $b$, described by the first term in Eq.~\ref{eq:2expdecay}. The cutoff frequency is given by
\begin{equation}
\nu_{\rm min}= \eta \, \frac{c}{\pi d} ,  
\label{eq:cutoff}
\end{equation}
where $c$ is the sound velocity, $d$ is the diameter of the crystal, and $\eta$ is a numerical constant that equals 2.05 for one spherical particle \cite{Lamb1881} and phonon modes with negligible broadening. According to this formula, YAG particles (where $c=6400$~m/s \cite{mezeix_comparison_2006}) with diameters below 100~nm should not be able to support phonons at $\Delta=35$~GHz, corrresponding to the splitting between the two crystal-field levels $^5{\rm D}_4 \, a/b$. Because our powder samples inevitably exhibit some distribution of particle sizes (the horizontal error bars in Fig. \ref{fig:Anr_Tb_vs_size} correspond to one standard deviation in the respective size distribution), the non-radiative relaxation should be suppressed in more and more particles as the mean of the size distribution is reduced below $\sim$~100~nm.  This should lead to a gradual decrease in the amplitude of the fast component $\alpha$ of the fluorescence decays. Note that increased inhomogeneous broadening causing partial overlap of the levels $^5{\rm D}_4 \, a$ and $b$ could also lead to similar observations. In our case, we account for this effect by computing the actual amplitude $\alpha$ from the measured one $\tilde{\alpha}$ as described above. As shown in Fig.~\ref{fig:Anr_Tb_vs_size}, we find that for particles created using method 1, the decay ratio begins to decrease at around 130~nm diameter, whereas the onset for particles created via method 2 starts at sizes of $\sim$~70 nm.

We modeled this effect by calculating the VDOS \cite{Lutz2016}, assuming that each mode contributes a Lorentzian with a width $\Delta\nu$ to the VDOS, for $10^5$ particles with different particle diameter following a Gaussian distribution. We obtained the mean diameter and the standard deviation of this distribution from the SEM images of our samples. We assumed that ions in a nanocrystal with diameter $d$ have a fast decay only if the VDOS of the nanocrystal at $\Delta = 35$~GHz is greater than zero. 
For each of the $10^5$ particles, we simulate the fluorescence decay, either a sum of two exponential functions in case the fast decay is allowed or only a single exponential decay with the radiative decay time in case the fast decay is forbidden. We then average the individual decays to obtain the overall fluorescence decay for the ensemble of $10^5$ particles. Finally, we fit it with a sum of two exponential  functions which results in the amplitudes $\alpha$ and $\beta$ of the fast and slow decay, respectively.
The only free parameter in the simulations is the width $\Delta\nu$ of the individual vibrational modes, which we obtain through a fit to the experimental data. Note, the mode width is forced to be the same for all powders fabricated using a specific method. The results of the simulations are presented in Fig.~\ref{fig:Anr_Tb_vs_size}. 

For large crystal diameters $d$, the simulated amplitude ratios (for crystals created via either method) are around 2, which corresponds to the bulk value at 1.5~K. When we reduce $d$, the amplitude ratios start decreasing at two distinct diameters (130 nm for method 1 and 70 nm for method 2 crystals). This can be explained by  different mode widths ($\Delta\nu=0.5$ and $6$~GHz for powders from methods 1 and 2, respectively) obtained from the simulations. If phonon modes are broad, as in powders from method 2, it is more likely that they overlap with the transition between the $^5{\rm D}_4 \, a/b$ levels, even for small crystals. However, for sharp modes (powders from method 1), overlap becomes significant only for larger particles in which more modes exist. Note that the width of the phonon modes is related to the powder quality. In particular, crystallites with reduced surface roughness should feature narrower phonon modes. This leads us to conclude that powder 1 should be of higher quality, which, however, cannot be verified given the insufficient resolution of our SEM pictures. Note that the optical inhomogeneous linewidths suggest that method 2 produces powders with less internal strain; however, it is not known if there is a relationship between particle surface morphology and internal strain.


Overall, the simulated values are in good agreement with the experimental data for powders produced by either method, consistent with suppression of phonon-induced relaxation in sufficiently small powders. In particular, we observed the complete transition from large particles, where the relevant phonon processes are fully allowed, to the smallest particles, where we could not measure any contribution of the phonon-induced component to population relaxation.

In addition to a change in the amplitude of the fast decay, we also expect a change in its characteristic decay time $T_\alpha$. For the fraction of particles with $\nu_{\rm min} \simeq \Delta_e$, the phonon density of states at $\Delta_e$ should deviate from the bulk value. With small enough particles we expect a decrease in VDOS as the lowest phonon mode moves towards higher frequencies. Thus, because the rate of the non-radiative relaxation is proportional to the VDOS, the phonon-induced decay rate for these nanocrystals should be slower than the one for the bulk, i.e. $T_\alpha$ should increase with decreasing particle size. Note that in some cases, an enhanced VDOS can occur due to phonon confinement, which would lead to faster decay rates. 
Experimentally (see inset Fig.~\ref{fig:Anr_Tb_vs_size}), we observe a decrease in $T_\alpha$ for some particles but we do not observe the expected increase for particles from method 1. This is consistent with the conjecture of having sharp phonon modes for particles from method 1 (resulting from the fit of the decay ratios in Fig.~\ref{fig:Anr_Tb_vs_size}), in which case the phonon cutoff occurs abruptly as the size is reduced so that the particles either experience a phonon rate equal or larger than the bulk or no phonon decay at all. 
For powders from method 2, the fit of the decay ratios predicts broader phonon modes, and we thus expect a smoother transition (rather than an abrupt cutoff as for particles from method 1)in the VDOS from the bulk value to zero. Therefore, more particles should experience decay times longer than the bulk before the decay is suppressed completely. Indeed, except for the smallest powder created using method 2, we see signatures of such an increase in $T_\alpha$, consistent with the predictions of our model. At this stage, the scatter of our experimental data unfortunately does not allow for a more in-depth analysis and interpretation.

\section{Measurement of temperature dependent population re-distribution}
Another indication of the restriction of non-radiative transition processes, including phonon modes, is the inhibition of thermalization (population re-distribution) between the two crystal-field levels $^5{\rm D}_4 \, a/b$ after their initial population through decay from $^5{\rm D}_4 \, f$. The ratio  $n_b/n_a$ will subsequently change and approach thermal equilibrium due to  any non-radiative transitions between $^5{\rm D}_4 \, a/b$. For $t >T_\alpha$, the population ratio is described by

\begin{equation}
n_b/n_a =(1-N_0) \, e^{-\Delta/k_B T}+N_0 
\label{eq:boltzmann}
\end{equation}
i.e. the Boltzmann distribution that assumes thermalization through non-radiative processes, with an additional offset $N_0$ that accounts for the overlap of the studied transitions, as described below. The population ratio $n_b/n_a$ for $t>T_\alpha$ is directly related to the ratio of the amplitudes $\beta_{b}$ and $\beta_{a}$ of the long fluorescence decay components (see Eq.~\ref{eq:2expdecay}): $\beta_{b}/\beta_{a}\propto~n_b(t)/n_a(t)$. This allows us to extract the actual population distribution for different temperatures. 

\begin{figure}[t!]
\centering
\includegraphics[width=1\columnwidth]{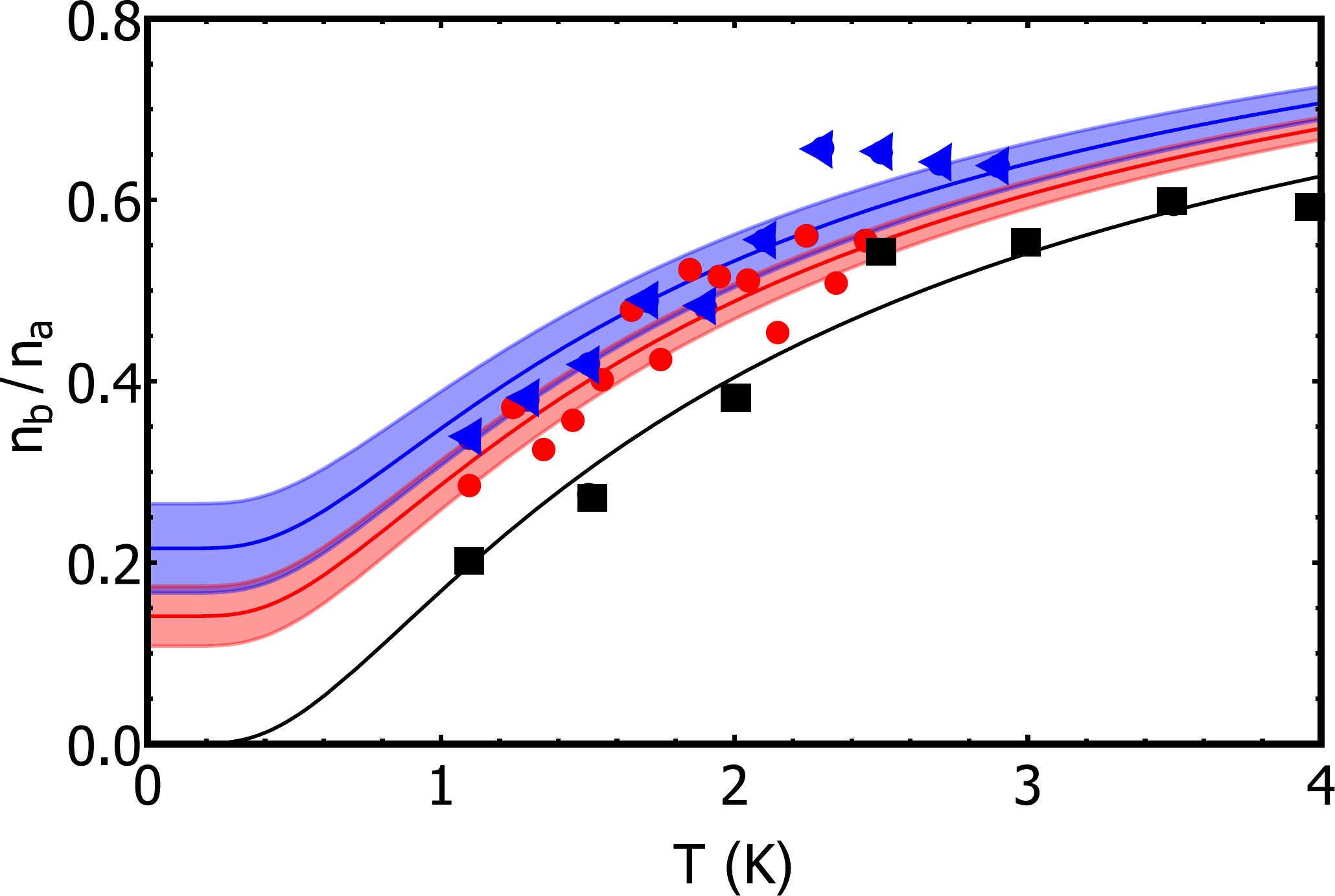}
\caption{Ratio of population $n_b/n_a$ as a function of temperature in the bulk crystal (black squares) and in nanocrystals from method 1 of average diameter $d=72$~nm (blue triangles) and 40 nm (red dots). Solid lines are best fits using Eq.~\ref{eq:boltzmann}, and shaded areas represent uncertainties.}
\label{fig:thermalization}
\end{figure}

The measured temperature dependence of $n_b/n_a$, calculated by averaging $\beta_{a}$ (and $\beta_{b}$) over both transitions starting in $^5$D$_4\, a$ (and $^5$D$_4\, b$), is shown in Fig.~\ref{fig:thermalization} for the bulk crystal as well as for two nanocrystal samples ($d=40$ and 72~nm) created via method 1. We fit the experimental points using Eq.~\ref{eq:boltzmann} after fixing the energy separation $\Delta_e$ to the bulk value of 35 GHz and leaving $N_0$ as a free variable. For both powders we observe a non-zero offset $N_0$. The fit gave $N_0$=0.21$\pm$0.1 and 0.14$\pm$0.07  for the two powders with $d=72$~nm and 40 nm, respectively, and for the bulk crystal it resulted in $N_0$=0, suggesting that there is the expected difference in thermalization in the powders. However, this offset can be fully explained by the amount of emission $P_{a,b}$ originating from the neighboring transition due to the increased inhomogeneous broadening as discussed above. Indeed, calculating $N_0$ by taking only line overlap into account, we find $N_0=$ 0.3$\pm$0.02 and 0.26$\pm$0.02 for $d=72$~nm and 40 nm, respectively. This seems to be at odds with the observed reduction of the nanocrystal's fast decay amplitude (Fig. \ref{fig:Anr_Tb_vs_size}), which is also supported by VDOS simulations. Thus, the observation of thermalization indicates that other processes, happening on time scales smaller than the 100 ns resolution of our detector, are responsible for population re-distribution in our smallest nanocrystals. 
Such fast relaxation could be caused by coupling of the Tb$^{3+}$ ions to tunneling modes characteristic of amorphous materials \cite{anderson_anomalous_1972,phillips_tunneling_1972} (note that the increase of amorphous character as the particle size is reduced is supported by the observation of larger inhomogeneous broadening). Other explanations are relaxations driven by  energy transfer \cite{Forster} or interactions between ions and surface states.

\section{Conclusion}
In conclusion, we observed modifications in relaxation dynamics between crystal field levels of \tbyag\  crystals as the particle size is varied from bulk to 40 nm, and confirmed via absorption measurements that this effect is not due to local heating. One possible explanation is a modification of the VDOS in the nanocrystals, which restricts phonon processes between the two first crystal-field levels in the $^5{\rm D}_4$ excited state that are separated by 35 GHz. 

However, other measurements suggest a different explanation: population redistribution is still observed within the two closely-spaced levels, meaning that other, fast, non-radiative processes must enable this transition. These processes may arise from a partially amorphous character of the nanocrystals, even though significant effort was dedicated to achieving good crystal quality by exploring various synthesis methods and modifying different important parameters in each of them, such as the addition of surfactants or the annealing temperature. We note that the case of YAG is particularly difficult because of the high annealing temperature---which favors particle growth--- required to crystallize the particles. 

Improving the nanocrystal quality by optimizing fabrication methods, as well as switching to a different material with a lower annealing temperature, such as fluoride crystals, may enable one to observe the full phonon restriction. However, there may exist a fundamental limit to how small a particle can become while still preserving the spectroscopic properties of a large crystal -- this limit is frequently estimated to be around 10 nm \cite{Bunzli}. Our results suggest that, for YAG, it may be around 100 nm. Measurements of crystal structure may shine more light on this important question.\\

\section{Acknowledgements}
The authors acknowledge support from Alberta Innovates Technology Futures (ATIF), the National Engineering and Research Council of Canada (NSERC), and the National Science Foundation of the USA (NSF) under award nos. PHY-1415628 and CHE-1416454, and the Montana Research and Economic Development Initiative. W. T. is a Senior Fellow of the Canadian Institute for Advanced Research (CIFAR).

\section{APPENDIX A: Powder characterization}

\begin{figure}[]
\centering
\includegraphics[width=1\columnwidth]{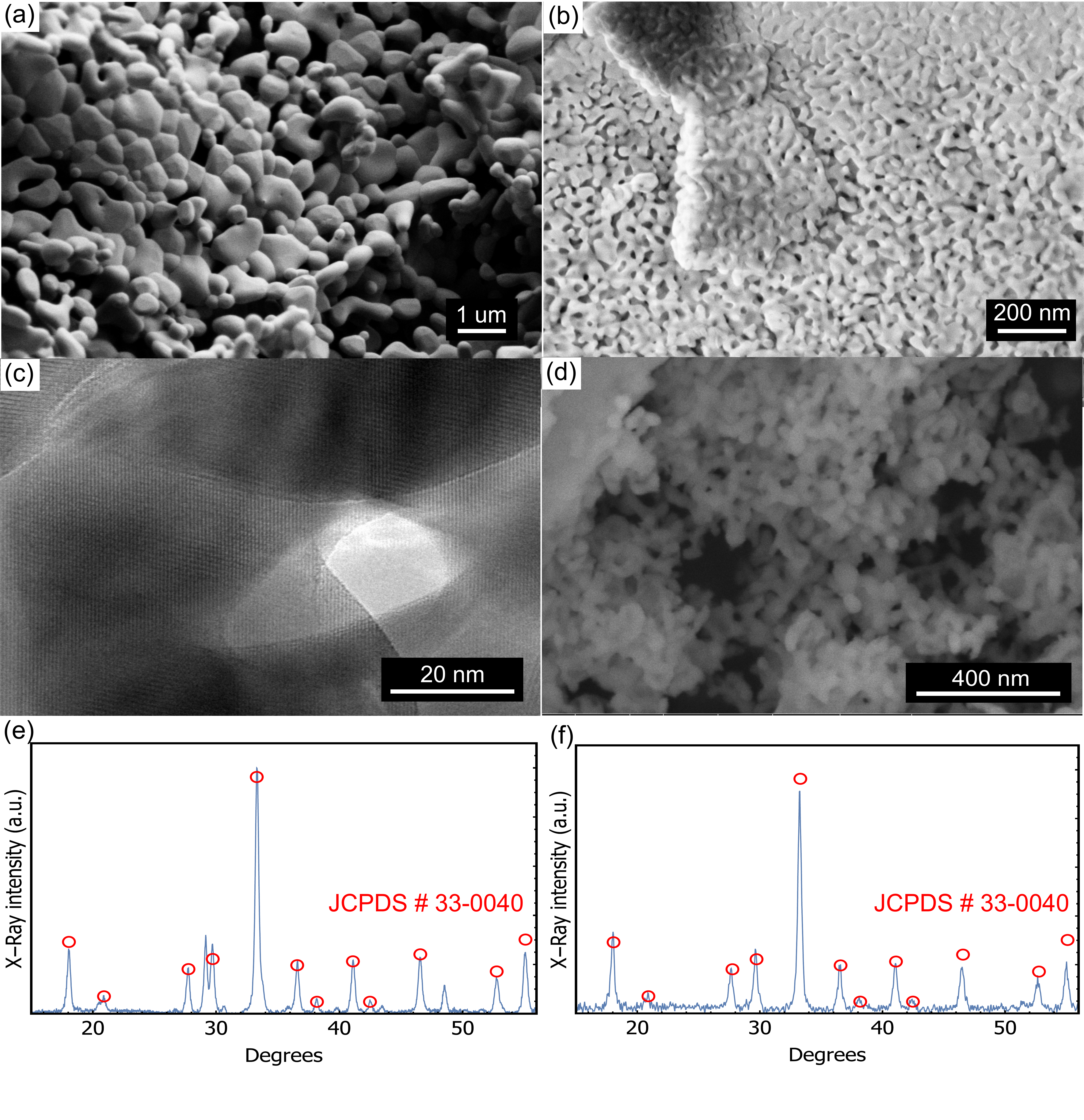}
\caption{Microscope images of 1\% \tbyag \ created using method 1 and annealed at 1400 C (a), or 900 C (b,c), and from method 2  annealed at 900 C (d). Panels (a,b,d) are SEM images, showing the size distribution of the nanocrystals. Panels (e) and (f) show XRD spectra of powders produced by method 1 and 2, respectively (solid blue lines), and the corresponding reference spectrum (JCPDS \# 30-0040; red circles) for YAG. Panel (c) is a high-resolution TEM image showing the crystalline structure (narrow white lines), which can extend over several particles if they are agglomerated.}
\label{fig:images}
\end{figure}

For all investigations of population dynamics presented in the main text, information about the morphology and size distribution of the various powders is needed to interpret the results. We obtained this information for powders from method 1 and 2 using a scanning electron microscope (SEM), with example images shown in  Fig. \ref{fig:images}. The images indicate that method 2 produces slightly less agglomerated powders compared to method 1. In addition, we confirmed that we obtained good single-phase crystalline YAG particles using powder x-ray diffraction (XRD) analysis. Figures \ref{fig:images} (e,f) show the perfect overlap between the XRD spectrum of the \tbyag\ powders produced by method 1 and 2 with the reference spectrum for YAG (JCPDS \# 30-0040). For selected powder samples, we also directly probed the quality of the crystal structure using a transmission electron microscope (TEM), as shown in  Fig. \ref{fig:images} (d). The TEM analysis revealed that the crystallite orientations in agglomerated nanocrystals can remain nearly aligned throughout multiple grains when they are fused together. Since it is possible that phonon modes extend across several crystallites in these cases, we considered the effective particle size in agglomerated samples to be equal to the larger size of the approximately aligned agglomerations rather than the individual grain size. The effect of agglomeration on phonon propagation dynamics  cannot be quantified at this stage.\\ 

In addition to using the TEM measurements, we also investigated the degree to which the crystallites within the powder act as isolated particles versus being part of a larger agglomerated mass by observing the increase in radiative lifetime as the particle sizes are reduced. As known from relations such as the Strickler-Berg equation \cite{Strickler62}, the radiative lifetime $T_{\rm rad}$ of an electric dipole transition depends on the average index of refraction, $n$, of the material.
Because of this, when the size of a fluorescing particle becomes comparable to the wavelength of the emitted light, the index of the medium surrounding the particle can have a significant effect on the radiative lifetime\cite{Yang1999,Schniepp02,Aubret16}. Consequently, we expect to observe an increase in the fluorescence lifetime as the average particle size in our powders is reduced. Note that strongly agglomerated particles would effectively  act as a single, larger particle in this case.\\ 

We employ a simple analytical model to estimate the size dependence of the fluorescence lifetime for perfectly isolated particles. More precisely, we used the form of the Lorentz local field, sometimes referred-to as the virtual cavity model \cite{Scheel1999}, where the radiative lifetime in the medium $T_{\rm rad}$ is related to the lifetime in vacuum $T_0$ according to $1/T_{\rm rad} = (1/T_0)  \, n_{\rm eff}(n_{\rm eff}^2+2)^2 / 9$, with $n_{\rm eff}$ being an effective index of refraction averaged over the surrounding medium within a distance on the order of the wavelength of light from the ion.  

For particles smaller than the wavelength of light, the electric field extends beyond the particle. To evaluate $n_{\rm eff}$ for such particles, we assumed that the electric field of the emitted light experiences the bulk crystal dielectric constant within the particle, and the vacuum dielectric constant outside the particle.  We furthermore assumed that the electric field outside the particle decays as $E(r)=E_0 e^{-\frac{r}{l}}$ over the decay length $l$, which is equal to the evanescent field decay length outside of a bulk dielectric given by $l=\lambda_0/2 \pi\sqrt{n^2-1}$ \cite{Fornel00}. Here $n$ is the refractive index of the bulk material and $\lambda_0$ the wavelength of the transition. The value of $n_{\rm eff}$ experienced by the emitting ion, and the resulting change in radiative lifetime, was then estimated by calculating the field-strength-weighted average dielectric constant over the area of non-zero  electric field.\\

\begin{figure}[t]
\centering
\includegraphics[width=1\columnwidth]{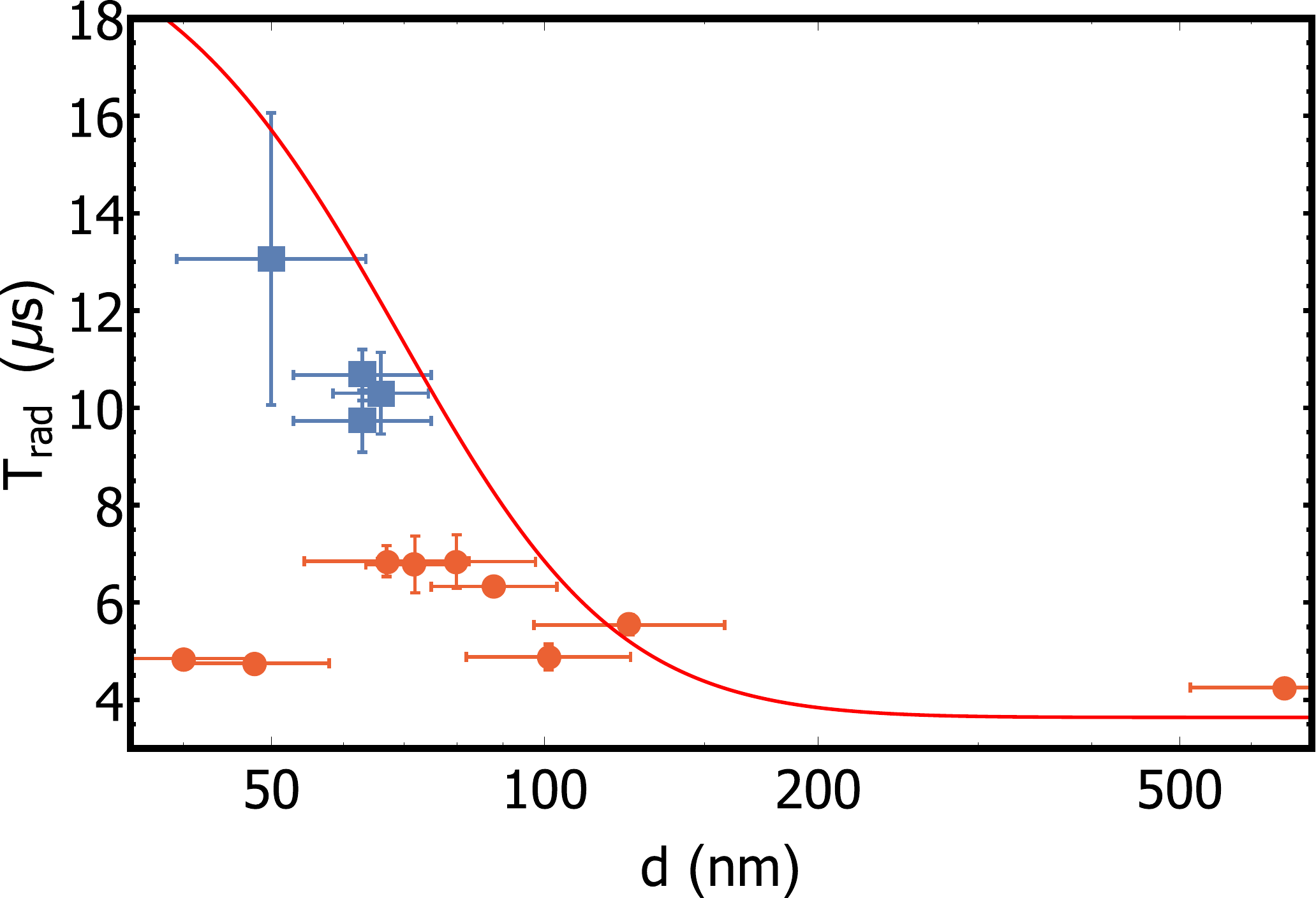}
\caption{Particle size ($d$) dependence of the radiative lifetime of the $^5$D$_4\, a$ level for powders from method 1 (red circles) and powders from method 2 (blue squares). The solid line shows the expected dependence.}
\label{fig:trad}
\end{figure}

By using this simple model, we estimated the change in lifetime with particle size using only the known bulk crystal index, transition wavelength, and lifetime with no free parameters, resulting in the solid line in Fig.~\ref{fig:trad}. We find that the measured lifetimes in our samples agree reasonably well with the calculated dependence (Fig.~\ref{fig:trad}). For method 1, the radiative lifetime increases up to 7 ms, as the crystallite size decreases, but powders smaller than 50 nm show lifetimes similar to the bulk crystal, indicating that some degree of agglomeration is present. For method 2, the radiative lifetime increases up to 13 ms, indicating that the particles in powders synthesized with this method indeed behave as individual particles with sizes approximately equal to the values estimated from the SEM and TEM analysis.

\section{APPENDIX B: Spectroscopic investigations of powder quality}

\begin{figure}[t]
\centering
\includegraphics[width=1\columnwidth]{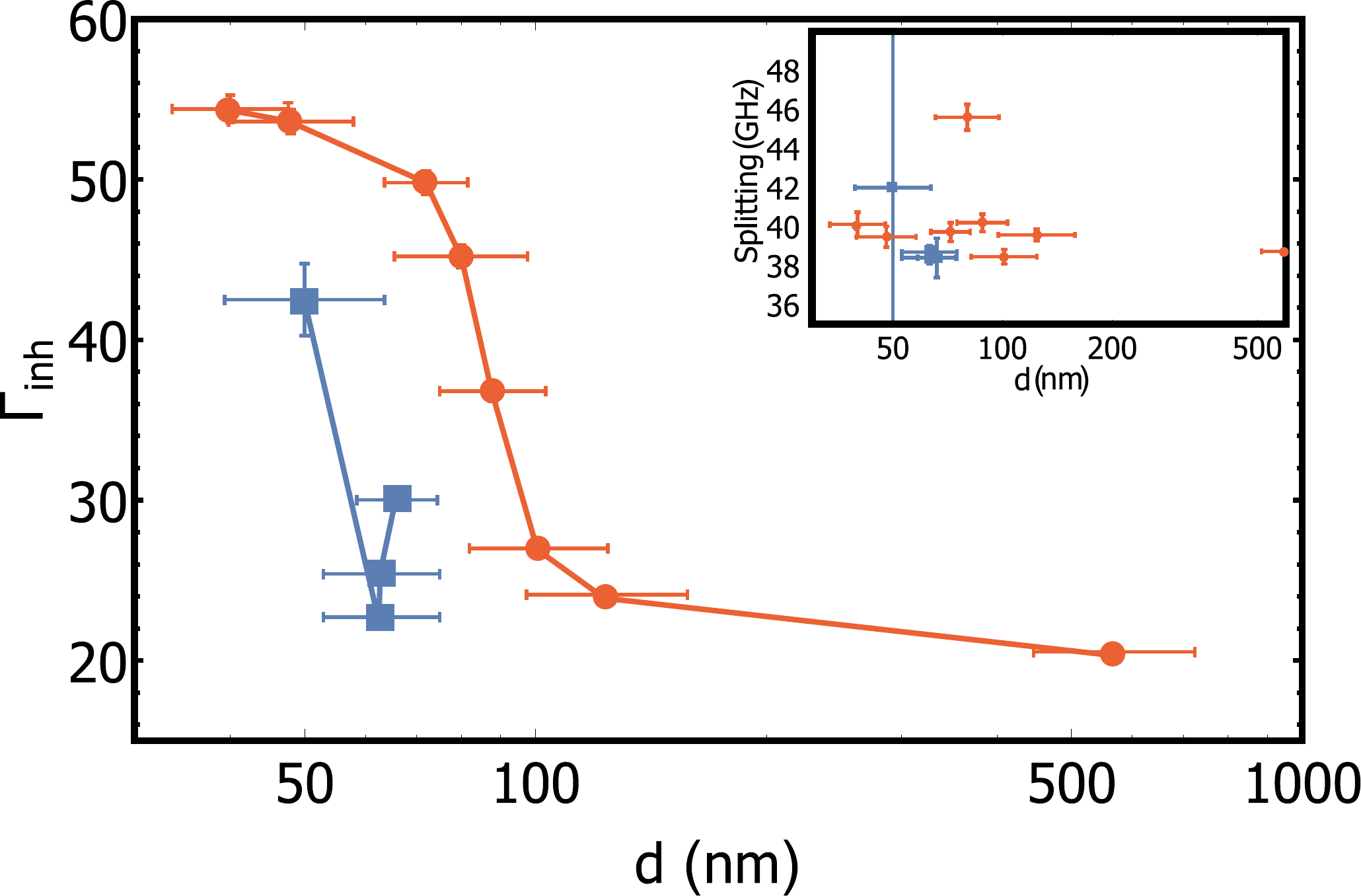}
\caption{Particle-size-dependence of the inhomogeneous linewidth for powders from method 1 (red circles) and powders from method 2 (blue squares). Inset:  Splitting $\Delta$ between the $^5$D$_4 \, a/b $ levels versus particle size. The solid lines are guides to the eye.}
\label{fig:inh}
\end{figure}

To ensure that we can selectively collect fluorescence from the two excited levels $^5$D$_4 \, a/b$ for each of our samples, we recorded fluorescence spectra by scanning the monochromator over the four lines connecting $^5$D$_4 \, a/b \rightarrow \, ^7$F$_5 \, a/b$. As shown in Fig.~\ref{fig:inh}, we observed that for both fabrication methods, the smallest nanocrystals feature an increased inhomogeneous broadening compared to the bulk. This increase, which was expected due to an increased amount of strain, was not observable in the XRD spectra due to the limited resolution of our XRD diffractometer (Rigaku Multiflex). The observation of increased inhomogeneous broadening is consistent with the emergence of relaxation that is facilitated by amorphous phases and surface defects (see main text). However, as shown in the inset of Fig.~\ref{fig:inh}, the splitting  $\Delta$ between the $^5$D$_4 \, a/b$ levels does not change with particle size, which indicates that the ions' crystal field splittings and local lattice symmetry are not measurably different in the small powders. 

\section{APPENDIX C: Local temperature measurement}

\begin{figure}[t]
\centering
\includegraphics[width=1\columnwidth]{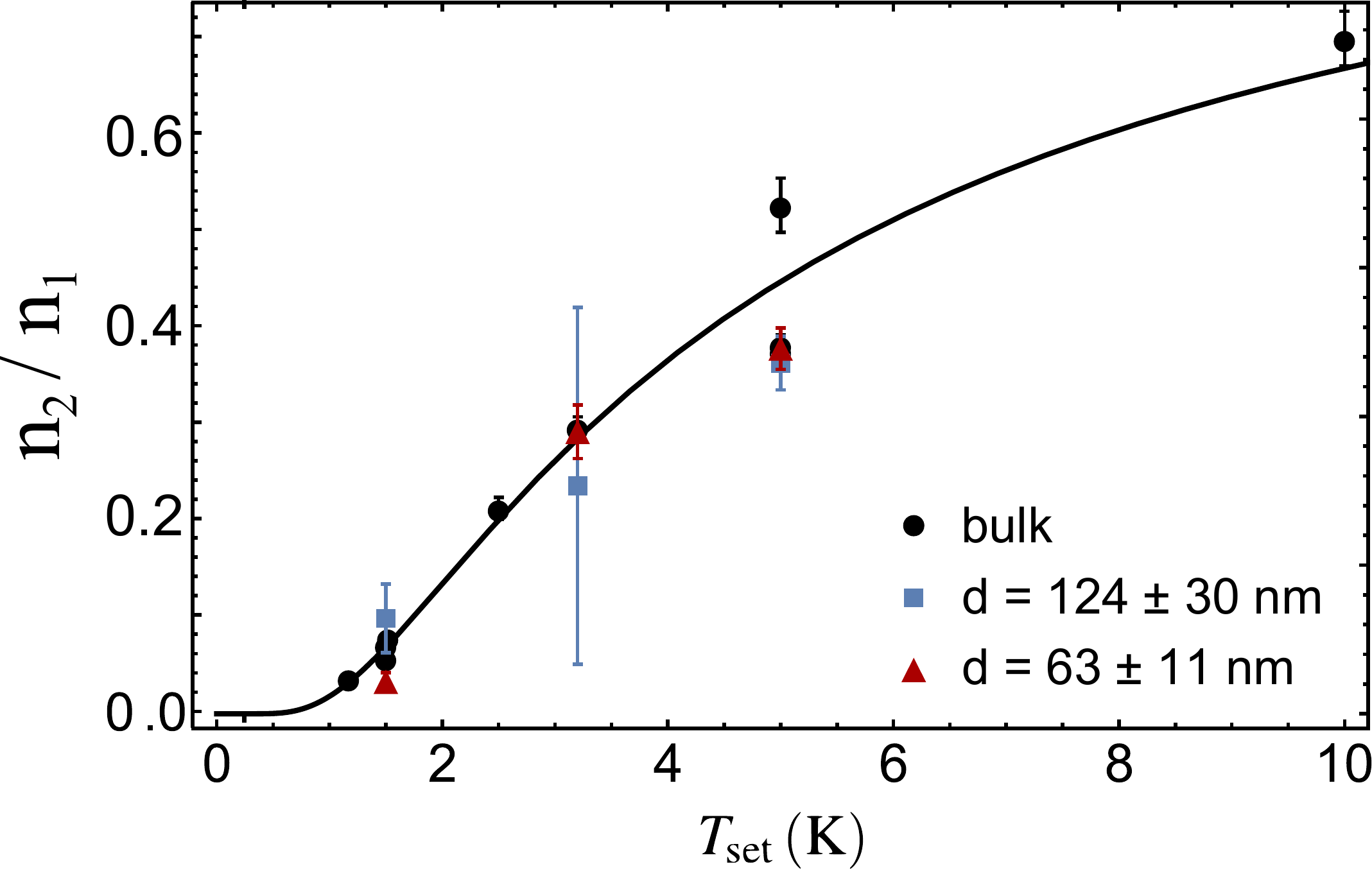}
\caption{Ratio $n_2/n_1$ of populations in the ground manifold $^7$F$_6 \; b/a$ levels as a function of the temperature $T_{\rm set}$ read by the cryostat sensor for a bulk crystal (black dots) and  nanocrystalline powder samples with diameters $d=124\pm 30$~nm and produced by method 1 (blue squares) and $d=63 \pm 11$~nm, produced by method 2 (red triangles). The solid line is the fit to a Boltzmann distribution with $\Delta_g = 83.5$~GHz. Note that the large error bar for 3.2 K and $d = 124$~nm is caused by a large uncertainty of the fit to that particular absorption spectrum.}
\label{fig:tempcal}
\end{figure}

In past measurements, the thermal conductivity of small powders in a gas environment was observed to decrease with particle size due to two effects. First, the phonon scattering length is reduced in small powders, and second, heat flow is hindered by the surface resistance of the small particles contained in the powder \cite{Garrett1974,Rettelbach1995,Brodie1965}. These effects could lead to a locally elevated powder temperature, especially when the powder is probed using a high power laser. An elevated sample temperature in turn would produce significant changes in population relaxation and thermalization that could potentially be misinterpreted as arising from other effects. To ensure that the laser excitation did not induce localized heating, a direct temperature measurement that enables the true internal temperature of the particles to be monitored is required.

We measured the internal temperature for each sample by recording the absorption spectrum $^7$F$_6 \, a,b$ $\rightarrow$ $^5$D$_4 \, f$ and comparing the populations $n_1$ and $n_2$ in the first and second crystal field level $^7$F$_6 \, a$ and $b$ of the ground state multiplet, which are separated by $\Delta_g = 83$~GHz. This allowed us to determine the effective internal sample temperature $T$ via the Boltzmann law $n_2/n_1 = A \exp\{-\Delta_g /k_B T\}$ with $k_B$ the Boltzmann constant. We obtained the coefficient $A$, which depends on the relative oscillator strengths of the transitions, from a calibration measurement using a 1\% Tb:YAG bulk crystal, that we assumed to thermalize quickly to the temperature $T_{\rm set}$ that was measured by the cryostat sensor. The temperature-dependent population ratio $n_2/n_1$ is shown in Fig.~\ref{fig:tempcal} for a selected set of nanocrystals. The close overlap between the bulk and powder results confirms that all powders thermalize as well as the bulk when immersed in liquid helium or vapor and that the laser is not measurably heating the nanocrystals, even for the smallest sizes. Since these measurements are conducted using the same laser power and focusing parameters as those used for all other measurements described in the main text, they confirm that the changes in dynamics that we observe in the powders are not caused by local heating. Also, with this method, we ensure that the ions that are studied in the relaxation dynamics measurements (as in the main text) are also the same ions that are used to measure the temperature.


\begin{thebibliography}{10}
\newcommand{\enquote}[1]{``#1''}

\bibitem{Riedmatten2015}
H.~de~Riedmatten and M.~Afzelius, \emph{Quantum Light Storage in Solid State
  Atomic Ensembles} (Springer International Publishing, Cham, 2015), pp.
  241--273.

\bibitem{heshami_quantum_2016}
K.~Heshami, D.~G. England, P.~C. Humphreys, P.~J. Bustard, V.~M. Acosta,
  J.~Nunn, and B.~J. Sussman, Journal of Modern Optics \textbf{0}, 1 (2016).

\bibitem{Sipahigil2016}
A.~Sipahigil, R.~E. Evans, D.~D. Sukachev, M.~J. Burek, J.~Borregaard, M.~K.
  Bhaskar, C.~T. Nguyen, J.~L. Pacheco, H.~A. Atikian, C.~Meuwly, R.~M.
  Camacho, F.~Jelezko, E.~Bielejec, H.~Park, M.~Lon{\v c}ar, and M.~D. Lukin,
  Science \textbf{354}, 847 (2016).

\bibitem{Hensen2015}
B.~Hensen, H.~Bernien, A.~E. Dreau, A.~Reiserer, N.~Kalb, M.~S. Blok,
  J.~Ruitenberg, R.~F.~L. Vermeulen, R.~N. Schouten, C.~Abellan, W.~Amaya,
  V.~Pruneri, M.~W. Mitchell, M.~Markham, D.~J. Twitchen, D.~Elkouss,
  S.~Wehner, T.~H. Taminiau, and R.~Hanson, Nature \textbf{526}, 682 (2015).

\bibitem{Hedges2010}
M.~P. Hedges, J.~J. Longdell, Y.~Li, and M.~J. Sellars, Nature \textbf{465},
  1052 (2010).

\bibitem{saglamyurekbroadband2011}
E.~Saglamyurek, N.~Sinclair, J.~Jin, J.~A. Slater, D.~Oblak, F.~Bussi\`eres,
  M.~George, R.~Ricken, W.~Sohler, and W.~Tittel, Nature \textbf{469}, 512
  (2011).

\bibitem{Maurer1283}
P.~C. Maurer, G.~Kucsko, C.~Latta, L.~Jiang, N.~Y. Yao, S.~D. Bennett,
  F.~Pastawski, D.~Hunger, N.~Chisholm, M.~Markham, D.~J. Twitchen, J.~I.
  Cirac, and M.~D. Lukin, Science \textbf{336}, 1283 (2012).

\bibitem{hong2013}
S.~Hong, M.~S. Grinolds, L.~M. Pham, D.~Le~Sage, L.~Luan, R.~L. Walsworth, and
  A.~Yacoby, MRS Bulletin \textbf{38}, 155–161 (2013).

\bibitem{Powell98}
R.~C. Powell, \emph{Physics of Solid-State Laser Materials} (Springer-Verlag
  New York, 1998).

\bibitem{kenyon_recent_2002}
A.~J. Kenyon, Progress in Quantum Electronics \textbf{26}, 225 (2002).

\bibitem{Justel98}
T.~J\"ustel, H.~Nikol, and C.~Ronda, Angewandte Chemie International Edition
  \textbf{37}, 3084 (1998).

\bibitem{Wang10}
F.~Wang, Y.~Han, C.~S. Lim, Y.~Lu, J.~Wang, J.~Xu, H.~Chen, C.~Zhang, M.~Hong,
  and X.~Liu, Nature \textbf{463}, 1061 (2010).

\bibitem{Downing1185}
E.~Downing, L.~Hesselink, J.~Ralston, and R.~Macfarlane, Science \textbf{273},
  1185 (1996).

\bibitem{Ruan2006}
X.~L. Ruan and M.~Kaviany, Phys. Rev. B \textbf{73}, 155422 (2006).

\bibitem{coprec}
T.~Lutz, L.~Veissier, C.~W. Thiel, P.~J.~T. Woodburn, R.~L. Cone, P.~E.
  Barclay, and W.~Tittel, Science and Technology of Advanced Materials
  \textbf{17}, 63 (2016).

\bibitem{Lutz2017}
T.~Lutz, L.~Veissier, C.~W. Thiel, P.~J. Woodburn, R.~L. Cone, P.~E. Barclay,
  and W.~Tittel, Journal of Luminescence pp. 2--12 (2017).

\bibitem{Zhong2015}
T.~Zhong, J.~M. Kindem, E.~Miyazono, and A.~Faraon, Nat Commun \textbf{6}
  (2015).

\bibitem{Ferrier2013}
A.~Ferrier, C.~W. Thiel, B.~Tumino, M.~O. Ramirez, L.~E. Baus\'a, R.~L. Cone,
  A.~Ikesue, and P.~Goldner, Phys. Rev. B \textbf{87}, 041102 (2013).

\bibitem{Lutz2016}
T.~Lutz, L.~Veissier, C.~W. Thiel, R.~L. Cone, P.~E. Barclay, and W.~Tittel,
  Phys. Rev. A \textbf{94}, 013801 (2016).

\bibitem{Yang1999}
H.-S. Yang, S.~P. Feofilov, D.~K. Williams, J.~C. Milora, B.~M. Tissue, R.~S.
  Meltzer, and W.~M. Dennis, Physica B: Condensed Matter \textbf{263}, 476
  (1999).

\bibitem{Yang1999a}
H.-S. Yang, K.~Hong, S.~Feofilov, B.~M. Tissue, R.~Meltzer, and W.~Dennis,
  Journal of Luminescence \textbf{83}, 139  (1999).

\bibitem{Liu2002}
G.~K. Liu, H.~Z. Zhuang, and X.~Y. Chen, Nano Letters \textbf{2}, 535 (2002).

\bibitem{Mercier2006}
B.~Mercier, C.~Dujardin, G.~Ledoux, C.~Louis, O.~Tillement, and P.~Perriat,
  Journal of Luminescence \textbf{119}, 224  (2006).

\bibitem{Forster}
T.~F\"orster, Annalen der Physik \textbf{437}, 55 (1948).

\bibitem{mezeix_comparison_2006}
L.~Mezeix and D.~J. Green, International Journal of Applied Ceramic Technology
  \textbf{3}, 166 (2006).

\bibitem{kaithwas}
N.~Kaithwas, M.~Dave, S.~Kar, S.~Verma, and K.~S. Bartwal, Crystal Research and
  Technology \textbf{45}, 1179 (2010).

\bibitem{freeze_dry}
H.~Gong, D.-Y. Tang, H.~Huang, and J.~Ma, Journal of the American Ceramic
  Society \textbf{92}, 812 (2009).

\bibitem{Haensch1972}
T.~W. H\"{a}nsch, Appl. Opt. \textbf{11}, 895 (1972).

\bibitem{Lamb1881}
H.~Lamb, Proceedings of the London Mathematical Society \textbf{s1-13}, 189
  (1881).

\bibitem{anderson_anomalous_1972}
P.~W. Anderson, B.~I. Halperin, and C.~M. Varma, Philosophical Magazine
  \textbf{25}, 1 (1972).

\bibitem{phillips_tunneling_1972}
W.~A. Phillips, Journal of Low Temperature Physics \textbf{7}, 351 (1972).

\bibitem{Bunzli}
X.~C. G.~Liu, \emph{Handbook on the Physics and Chemistry of Rare Earths}
  (ELSEVIER SCIENCE \& TECHNOLOGY, 2007).

\bibitem{Strickler62}
S.~J. Strickler and R.~A. Berg, The Journal of Chemical Physics \textbf{37},
  814 (1962).

\bibitem{Schniepp02}
H.~Schniepp and V.~Sandoghdar, Phys. Rev. Lett. \textbf{89}, 257403 (2002).

\bibitem{Aubret16}
A.~Aubret, A.~Pillonnet, J.~Houel, C.~Dujardin, and F.~Kulzer, Nanoscale
  \textbf{8}, 2317 (2016).

\bibitem{Scheel1999}
S.~Scheel, L.~Kn\"oll, and D.-G. Welsch, Phys. Rev. A \textbf{60}, 4094 (1999).

\bibitem{Fornel00}
F.~de~Fornel, \emph{Evanescent Waves} (Springer-Verlag Berlin Heidelberg,
  2001).

\bibitem{Garrett1974}
K.~W. Garrett and H.~M. Rosenberg, Journal of Physics D: Applied Physics
  \textbf{7}, 1247 (1974).

\bibitem{Rettelbach1995}
T.~Rettelbach, J.~S\"auberlich, S.~Korder, and J.~Fricke, Journal of
  Non-Crystalline Solids \textbf{186}, 278  (1995).

\bibitem{Brodie1965}
D.~E. Brodie and C.~F. Mate, Canadian Journal of Physics \textbf{43}, 2344
  (1965).

\end{thebibliography}
\end{document}